# Free Energy of Activation for the Comorosan Effect

George E. Bass[a*], Bernd Meibohm[a], James T. Dalton[b] and Robert Sayre[c]

[a] College of Pharmacy, University of Tennessee Health Science Center, 800 Madison Avenue, Memphis, TN 38163, U.S.A.  [b] College of Pharmacy, Ohio State University, 500 West 12th Avenue, Columbus OH  43210, U.S.A.  [c] Rapid Precision Testing Laboratory, 8225 Rockcreek Parkway, Memphis, TN , 38018, U.S.A.

[*] Author to whom correspondence should be addressed. Email gbass@utmem.edu

Reaction rate data for lactic dehydrogenase / pyruvate, lactic dehydrogenase / lactate and malic dehydrogenase / malate enzyme reactions were analyzed to obtain activation free energy changes of –329, -195 and –221 cal/mole, respectively, for rate increases associated with time-specific irradiation of the crystalline substrates prior to dissolution and incorporation in the reaction solutions.  These energies, presumably, correspond to conformational or vibrational changes in the reactants or the activated complex.  For the lactic dehydrogenase / pyruvate reaction, it is estimated that on the order of 10% of the irradiation energy (546 ± 50 nanometers, 400 footcandles for 5 seconds) would be required to produce the observed reaction rate increase if a presumed photoproduct is consumed stoichiometrically with the pyruvate substrate.  These findings are consistent with the proposition that the observed reaction rate enhancements involve photoproducts of oscillatory atmospheric gas reactions at the crystalline enzyme substrate surfaces rather than photo-excitations of the substrate molecules, *per se*.

## 1. Introduction

Transition-state theory and its associated free energy of activation serve as the primary tools in conceptualization of models for enzyme catalysis mechanisms. These models are becoming increasingly elaborate, and hypothetical, as they struggle to account for the enzyme's specificity and efficiency (e.g., see Ma *et al*., [1]). Modeling necessarily is based on the assumption that all relevant features of the catalytic process are taken into consideration. With regard to this point, a body of experimental work referred to as the Comorosan effect may prove relevant. Here, we present an evaluation of some implications of that work from a transition-state theory perspective.

The Comorosan effect is a phenomenon in which the initial velocity of an enzymatic reaction is increased as a consequence of utilizing substrate that had been irradiated in the crystalline state, for a specific time duration, prior to dissolution and incorporation in the reaction mixture. This behavior has been observed for the reactions of over twenty enzymes isolated from multiple sources (see Table 1) and, thus, may reflect a very common, perhaps even fundamental, property of enzyme catalysis. To date, it has not been established how the relevant irradiation energy is absorbed by the crystalline material, how it is transformed on dissolution, nor how it is manifest in producing an enhanced *in vitro* enzymatic reaction rate. No assessment of the energetics attendant to the observed reaction rate stimulation has been reported. For overviews of much of the published work in this area, see Comorosan *et al* , [2, 3].

Comorosan [4,5] sought to explain the phenomenon as due to photo-excitation of the irradiated crystalline enzyme substrate molecules, *per se*, to special "biological observable" quantum states detectable only with the extreme sensitivity of enzymes. The purpose of this theoretical investigation is to derive an estimate of the magnitude of the energy that is involved, particularly with respect to Comorosan's model. We applied the transition-state theory of reaction rates to kinetic data published by Comorosan and co-workers for three enzyme reactions, the lactic dehydrogenase interconversions of pyruvate and lactate [6] and the malic dehydrogenase conversion of malate to α-oxaloacetate [7].

In the simplest representation of transition-state theory [8] for an enzymatic reaction, one has:

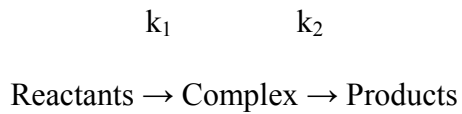

$$\text{Reactants} \xrightarrow{k_1} \text{Complex} \xrightarrow{k_2} \text{Products}$$

where $k_1$ is the rate constant describing formation of the activated complex and $k_2$ that for its conversion to product. Formation of the transition-state complex requires a free energy of activation, $\Delta G^*$, usually illustrated as a barrier along the reaction coordinate. The basic modeling assumption made here is that the energy absorbed and transferred from the irradiated crystals, in whatever form, serves to reduce $\Delta G^*$, thereby increasing the reaction rate. In general, this might be achieved either by raising the free energy level of the free reactants or of the Michaelis-Menten complex (if considered distinct from the transition-state complex), or by lowering that of the transition-state complex.

The rate, v, of the reaction can be expressed as:

$$v = [\text{Complex}] \times (\text{rate of traversing the energy barrier}).$$

The rate of traversing the energy barrier is given by $\kappa K_B T/h$, where $\kappa$ is a transmission coefficient giving the probability that formation of the complex will lead to reaction, $K_B$ is Boltzmann's constant, T is the absolute temperature and h is Planck's constant. The transmission coefficient usually is assumed to be unity or very nearly so, and will be taken as such here. Thus,

$$v = K_B T/h \, [\text{Complex}].$$

The equilibrium constant for the complex, $K^*$, is given by:

$$K^* = [\text{Complex}] / [\text{Reactants}].$$

Thus,

$$v = K^* \, K_B T/h \, [\text{Reactants}].$$

From thermodynamics one has: $-\Delta G^* = RT \ln(K^*)$, or, correspondingly,

$$K^* = \exp(-\Delta G^*/RT).$$

Thus,

$$v = (K_B T/h) \exp(-\Delta G^*/RT) \, [\text{Reactants}].$$

On the other hand, one may write:

$$v = k_1 \, [\text{Reactants}].$$

Equating the two expressions for v gives:

$$k_1 = (K_B T/h) \exp(-\Delta G^*/RT).$$

Or,

$$-\Delta G^* = RT \ln(k_1/\alpha), \text{ where } \alpha = (K_B T/h).$$

Consider the conversion of pyruvate to lactate catalyzed by lactic dehydrogenase (LDH), a reaction which includes nicotinamide adenine dinucleotide, reduced (NADH) as a cofactor. The reaction may be represented as:

Pyruvate + NADH + LDH → Lactate + $NAD^+$ + LDH .

Experimentally, the reaction rate can be assessed spectrophotometrically by recording the disappearance of NADH at 340 nm. The rate of formation of the products, lactate and $NAD^+$, is equal to the rate of loss of reactants, pyruvate and NADH. The rate of loss of pyruvate is equal to that of NADH, which is represented by decreasing absorbance of the reaction solution at 340 nm. Correspondingly, the LDH / lactate and MDH / malate reactions can be followed by the conversion of $NAD^+$ to NADH (increasing absorbance at 340 nm).

Of interest here is the limiting (maximal) rate of reaction, that is, the initial reaction rate, achieved when all reactants are present in excess relative to the enzyme. This is the effective rate at an instant after t = 0 and before a significant portion of any reactant can be removed or product accumulated.

## 2. Methods

In modeling the experimental data at hand, we assumed that irradiation of the crystalline substrate creates an entity, or precursor thereof, which though unidentified, we shall designate as Є. Inspection of the published reaction rate curves strongly suggest that it is consumed in the course of the reaction, the presumed product being here designated Є'.

This entity may, or may not, be identifiable with excited molecules of the substrate. In the subsequent enzyme assay, two simultaneous reactions can proceed:

(1) $S + E \leftrightarrows S{:}E \rightarrow P + E$

and

(2) $Ɛ + S + E \leftrightarrows Ɛ{:}S{:}E \rightarrow P + E + Ɛ'$

(for notational simplicity, the cofactor, NADH or $NAD^+$, is not represented).

Here, two competing models may be envisioned. If $Ɛ$ corresponds simply to an excited state of the substrate, as proposed by Comorosan, then the activation energy barrier for the catalyzed reaction will be reduced as a consequence of increased initial energy of the substrate. In this case, one would anticipate that the concentration of $Ɛ$ must be some appreciable fraction of that of the substrate. On the other hand, $Ɛ$ may correspond to an altogether different chemical species that interacts with the enzyme to alter its conformation in a manner that increases its catalytic potency, thus lowering the peak of the activation energy barrier. In this alternate case, the concentration of $Ɛ$ would need be only a fraction of the concentration of the enzyme, typically orders of magnitude less than that of the substrate.

Under the conditions of the experiments being examined, concentration of the reactants, other than $Ɛ$, are sufficiently high to be treated as constant. For the control reaction (non-irradiated substrate), the measured reaction rate is zero-order throughout with respect to both substrate and cofactor (e.g., pyruvate and NADH). Its rate constant is designated $k_0$. For the reaction involving irradiated substrate, the initial reaction rate is faster and then decreases (presumed due to consumption or degradation of $Ɛ$) to the same constant rate observed for the control reaction. Its initial rate constant at t=0 is designated $k_\varepsilon$.

Data points for NADH absorbance versus time were estimated visually from the published graphs for the three reactions by photo enlargement onto a grid (estimation error <2%). The two data sets for each reaction with and without prior irradiation (absorbance designated $A_\varepsilon$ and $A_0$, respectively) were utilized to determine $k_0$ and $k_\varepsilon$ by simultaneously fitting them to the following differential equations using nonlinear regression analysis:

*LDH/Pyruvate reaction:*

$dA_0/dt = -k_0$

$dA_\varepsilon/dt = -k_0 (1+\beta \exp(-k't))$

*LDH/Lactate and MDH/Malate reaction:*

$dA_0/dt = k_0$

$dA_\varepsilon/dt = k_0 (1+\beta \exp(-k't))$

Curve fit values for $k_\varepsilon$ and $k_0$ were converted from absorbance units / sec to moles NADH / sec noting that for the LDH / pyruvate reaction, the initial absorbance was 0.480 units for the solution containing 0.34 μmoles of NADH.

All calculations were performed using the software package Scientist V.2.01, MicroMath Inc., Salt Lake City, UT.

For reaction (1),

$$\Delta G^*(1) = -RT \ln(k_0/\alpha) \quad \text{(where } \alpha = K_B T/h),$$

and for reaction (2),

$$\Delta G^*(2) = -RT \ln(k_\varepsilon /\alpha),$$

so that,

$$\Delta\Delta G^* = \Delta G^*(2) - \Delta G^*(1) = -RT \ln(k_\varepsilon / k_0)$$

$$= -594 \ln(k_\varepsilon / k_0) \text{ cal/mole for } R = 1.987 \text{ cal/deg/mole and } T = 299 \text{ K}.$$

### 3. Results

The fitted curves for the three reactions are presented in Fig. 1, 2 and 3. The experimental parameters and calculated reductions in activation free energies, $\Delta\Delta G^*$, are presented in Table 2.

The LDH / pyruvate data were analyzed further to provide an estimate of the minimum amount of relevant energy that would have had to be absorbed in the irradiation step, assuming that $\epsilon$ corresponds to photo-excited substrate molecules. The 3 mL reaction mixture contained $2.2 \times 10^{-6}$ moles of pyruvate and $0.025 \times 10^{-6}$ g of LDH, estimated to correspond to $1.8 \times 10^{-11}$ moles of LDH (using MW = 140,000 daltons [9]). Inspection of the absorbance *versus* time data for this reaction revealed that the irradiation effect was associated with a decreased absorbance of 0.030 units that occurs entirely within the first 20 sec of the reaction. Assuming $\epsilon$ corresponds to photo-activated pyruvate molecules and that 100% of these were converted to lactate, then $2.1 \times 10^{-8}$ moles of such molecules would have been introduced into the reaction mixture. For $\Delta\Delta G^* = 329$ cal/mole, this implies that the total amount of associated energy transferred into the reaction cuvette was, minimally, $7.0 \times 10^{-6}$ cal. Since one-tenth of the solution of the irradiated crystals was transferred to the reaction mixture (S. Comorosan, personal communication), the dissolved crystals would thus have possessed on the order of $7 \times 10^{-5}$ cal of transferable relevant energy. (Data for the slower lactate and malate reactions did not lend themselves to a corresponding assessment.)

No direct measurements of the amount of irradiation actually absorbed by the crystals have been reported. In the two reports from which the kinetics data utilized here were taken, the irradiation intensity was characterized as being 600 lux. However, in a later publication noting more detailed study of this aspect, Comorosan indicated that the required irradiation intensity, the illuminance, should be approximately 400 footcandles [2], corresponding to on the order of 4000 lux. Using this latter figure (assumed more reliable) along with an estimate of 1 cm$^2$ for the area covered by the sodium pyruvate crystals and an exposure time of 5 seconds, the total energy to which the crystals were exposed is calculated to be approximately $7.4 \times 10^{-4}$ cal. This then would imply that approximately 10% of the total incident radiation must be captured such as to produce photo-excited substrate molecules.

**4. Discussion**

On the face of it, one would not expect the crystal irradiation procedure employed in the studies addressed here to have any measurable impact on a subsequent enzyme reaction rate. However, Comorosan and co-workers have published 15 primarily experimental reports involving over 20 different enzymes employing a wide range of reaction rate determination methodologies all of which display this behavior [10, 11,12,13]. In addition, 4 collaborative studies [14, 15, 16, 17] conducted in other laboratories and 3 independent studies [18, 19, 20] have been reported. Some of these include single and double-blind procedures to mitigate against the possibility of experimenter and procedural bias. While a trivial explanation may underlie the phenomenon, such has not been uncovered to date. The particular data examined here for LDH / pyruvate are consistent with other of the above cited reports for that particular reaction, as well as the body of work as a whole.

Early explanations for the phenomenon assumed that the irradiation procedure placed the substrate molecule (e.g., pyruvate, malate, etc.) in an excited state that could be detected (discerned) only by the extreme sensitivity of its enzyme. This model would imply that the photo-activated entity, here designated $\epsilon$, is, in fact, a sub-population of the substrate and would be consumed stoichiometrically in the conversion of substrate to product. However, it is estimated above that at least 10% of the irradiation incident on the crystalline material would need be absorbed to provide the required activation energy reduction. This seems highly unlikely, and possibly suggestive of a mysterious light-matter-biological interaction that lies completely outside contemporary scientific paradigms. More recently, an alternate model has been proposed [21] wherein the crystal irradiation process induces oscillatory free radical mediated reactions involving atmospheric gases at the surface of the crystals. This putative photo-driven process would be similar to a number of observed temperature-driven oscillatory systems involving atmospheric gases [22, 23, 24]. On cessation of irradiation, defining the t* period, much slower dark reactions would lead to relatively stable, water soluble chemical species which, in turn, are capable of altering reactivity of a particular enzyme. Thus, a much smaller quantity of the $€$ species would be required, perhaps only some fraction of the enzyme molecular concentration rather than the orders of magnitude higher substrate concentration (here, for the LDH reaction, $1.8 \times 10^{-11}$ *vs* $2.2 \times 10^{-6}$ moles/3 mL).. This behavior would be similar to that observed for the well known action of nitric oxide, an atmospheric gas photo-product, which can induce over a 40-fold increase in the rate of conversion of GTP to cAMP by guanylate cyclase [25, 26]. The body of work on this phenomenon reveals that different enzyme reactions may be enhanced by different crystal irradiation time periods (e.g., 15 s rather than 5 s exposure). Accordingly, it should be anticipated that a small set of relevant chemical species are generated in the photochemical and follow-on dark reactions.

Should this model be proved correct and if the phenomenon indeed reflects a universal property of enzymes, a new door might be opened to link known physico-chemical processes to pre-biotic evolution and the generalized non-linear dynamics espoused by others [27, 28] as fundamental to life processes. Moreover, at some level of refinement, transition state modeling of enzyme reactivity will need to take such a ubiquitous property into consideration.

Table 1. Species, enzymes and substrates for which phenomenon has been observed.

| Species | Enzymes | Irradiated Crystalline Substrates and Chemicals |
|---|---|---|
| *Bacillus subtilis*<br>*Bacillus cereus*<br>*E. coli*<br>*Salmonella panama*[#]<br><br>*Saccharomyces cerevisiae* (yeast)<br><br>*Canavalia ensiformis* (jack bean)<br><br>rat<br>rabbit<br>pig<br>beef<br>human[#]<br>chicken[#] | Aldolase<br>Citrate Synthase<br>Fumarase<br>Fructose-1,6-Diphosphatase<br>Glucose Dehydrogenase<br>Glucose-6-Phosphatase<br>Glucose-6-Phosphate Dehydrogenase<br>Glutamic Dehydrogenase<br>Glutamic-Oxalacetic Transaminase<br>Glutamic-Pyruvic Transaminase<br>Hexokinase<br>Invertase<br>Isocitrate Dehydrogenase<br>Lactate Dehydrogenase<br>Malate Dehydrogenase<br>Malic Enzyme<br>Penicillinase<br>Phosphoglucomutase<br>Phosphoglucose Isomerase<br>Phosphohexose Isomerase<br>Pyruvate Dehydrogenase<br>Succinate Dehydrogenase<br>Urease<br>Xanthine Oxidase | acetyl-Co-A<br>adenine<br>alanine<br>arginine<br>aspartate sodium<br>chloramphenicol hemisuccinate<br>cytidine<br>cytochrome C<br>fructose 1,6-diphosphate sodium<br>glucose<br>glucose 1-phosphate<br>glucose 6-phosphate<br>glutamate sodium<br>histidine<br>isocitrate sodium<br>α-ketoglutarate potassium<br>lactate lithium<br>malate sodium<br>6-mercaptopurine<br>mitomycin C<br>oxaloacetate<br>penicillin sodium<br>potassium chloride[#]<br>pyruvate sodium<br>silicon dioxide[#]<br>sodium bromide[#]<br>sodium chloride<br>succinate sodium<br>sucrose<br>tetracycline HCl<br>thymine<br>tryptophan<br>urea<br>xanthine sodium |

[#] unpublished, Bass et al.

Table 2. Reaction Rate Parameters

| Reaction enzyme/substrate/cofactor | Substrate μmoles/3 ml | Cofactor μmoles/3 ml | Enzyme U/3 ml | $t^*$, sec | $k_\varepsilon$, $10^{-9}$ | $k_0$, $10^{-9}$ | $\Delta\Delta G^*$, cal/mole | Ref |
|---|---|---|---|---|---|---|---|---|
| LDH/Na pyruvate/NADH | 2.20 | 0.34 | $9 \times 10^{-3}$ | 5 | 4.95 | 2.84 | -329 | a |
| LDH/Li lactate/NAD | 4.10 | 3.0 | $18 \times 10^{-3}$ | 15 | 4.51 | 3.25 | -195 | a |
| MDH/Na malate/NAD | 64 | 3.0 | $450 \times 10^{-3}$ | 25 | 35.7 | 24.6 | -221 | b |

Temperature = 299 K for each reaction.
$k_0$ = initial reaction rate constant obtained using non-irradiated substrate, moles NADH / sec.
$k_\varepsilon$ = initial reaction rate constant obtained using irradiated substrate, moles NADH / sec.
$t^*$ = duration of crystalline substrate irradiation at 546 nm.
$\Delta\Delta G^* = -594 \ln(k_\varepsilon / k_0)$ cal/mole.
References: a. S. Comorosan et al., 1972; b. S. Comorosan et al., 1971c.
LDH = lactic dehydrogenase; MDH = malic dehydrogenase; NAD = nicotinamide dinucleotide; NADH = nicotinamide dinucleotide, reduced.

Fig. 1. Computed Curves for the Lactic Dehydrogenase / Pyruvate Reaction. Data points estimated from Comorosan et al., 1972, Figure 3A. **George E. Bass, Bernd Meibohm, James T. Dalton and Robert Sayre**

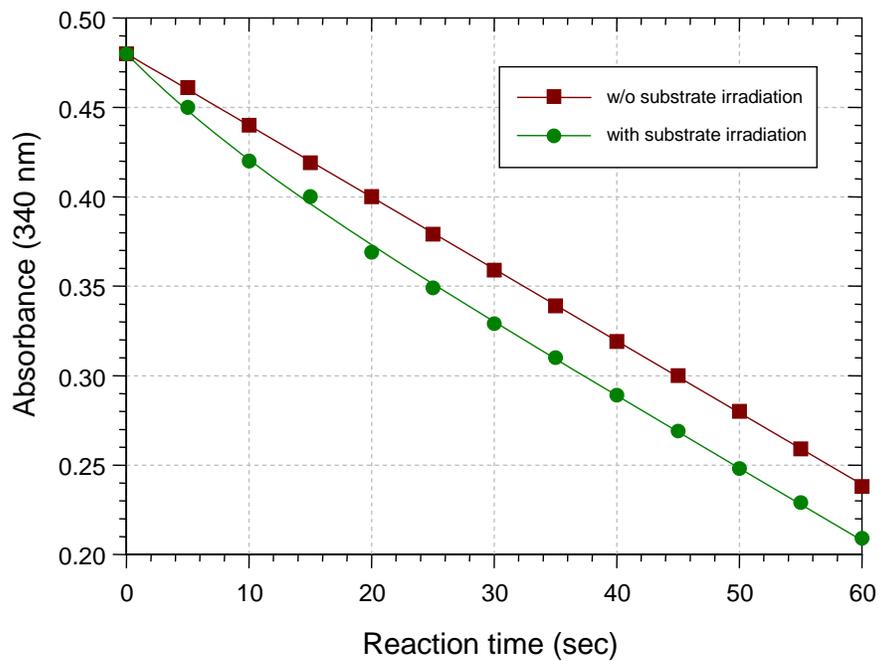

Fig. 2. Computed Curves for the Lactic Dehydrogenase / Lactate Reaction. Data points estimated from Comorosan et al., 1972, Figure 3B. **George E. Bass, Bernd Meibohm, James T. Dalton and Robert Sayre**

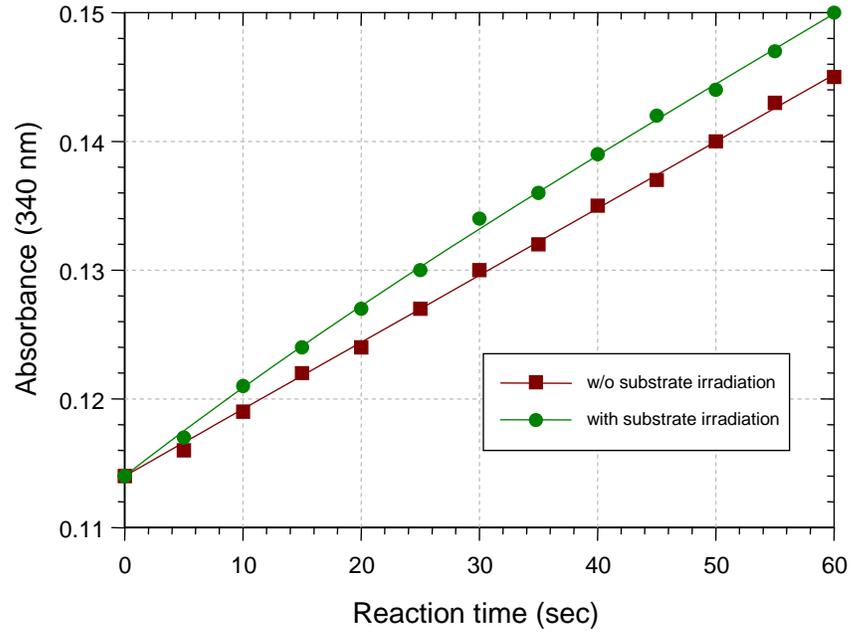

Fig. 3. Computed Curves for the Malic Dehydrogenase / Malate Reaction. Data points estimated from Comorosan et al., 1971c, Figure 1B., **George E. Bass, Bernd Meibohm, James T. Dalton and Robert Sayre**

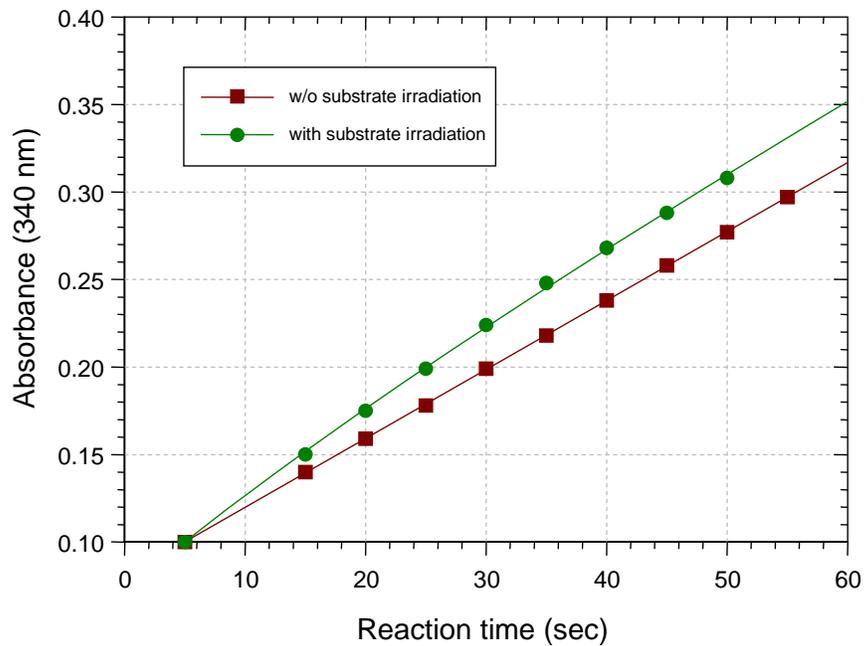

**Acknowledgement**